\documentclass[preprintnumbers,amsmath,amssymb,prd,twocolumn,nofootinbib,showpacs]{revtex4}
\usepackage{amsmath,amssymb,graphicx,paralist,comment}

\begin{document}

\title{Comment on ``Modified scalar-tensor-vector gravity theory and the constraint on its parameters'' by Deng, et al.}

\author{J. W. Moffat$^{\dag *}$ and V. T. Toth$^{\dag}$\\~\\}

\affiliation{$^\dag$Perimeter Institute, 31 Caroline St North, Waterloo, Ontario N2L 2Y5, Canada}
\affiliation{$^*$Department of Physics, University of Waterloo, Waterloo, Ontario N2L 3G1, Canada}

\begin{abstract}
We comment on a recent paper \cite{Deng2009} by Deng et al. in which the Eddington-Robertson parameters for our modified gravity theory (MOG) are derived. We show by explicit calculation that the role of the vector field $\phi_\mu$ cannot be ignored in this derivation.
\end{abstract}


\maketitle

In \cite{Deng2009}, Deng et al. present a generalization of STVG \cite{Moffat2006a,Moffat2007e}, a scalar-tensor-vector formulation of our modified gravity theory (MOG). Their argument is based on a Parameterized Post-Newtonian (PPN, \cite{Chandra1965,Will1993}) representation of STVG, which they derive in the appendices of their paper. In this derivation, the authors explicitly ignore the MOG vector field: ``Then, we rewrite Eqs. (35) and (42) while abandoning the vector field [...]''. Through this approach, the authors reduce MOG to a specific case of Brans-Dicke theory \cite{BD1962}, depriving MOG of the essential feature of a repulsive massive vector field.

Yet the vector field cannot be ignored. Considering Eq.~(\ref{eq:12}) in \cite{Deng2009}, we find that the term $\omega m^2k^2\phi_\mu\phi_\nu/c^2$ appears in sums also containing the trace of the matter stress-energy tensor. Clearly, this term can be omitted only if the condition $\omega m^2k^2\phi_\mu\phi^\mu/c^2\ll T$ is satisfied. Otherwise, this term must be retained. Considering our spherically symmetric, static vacuum solutions, this is precisely the case: for instance, in the near vacuum of the outer solar system, $T$ is vanishingly small, yet $\phi_0$ is nonvanishing due to the ``fifth force'' influence of the Sun. Even near the Sun, for instance along the path of a radio signal received from distant spacecraft at opposition, the energy density associated with the fifth force vector field is comparable to the energy density of the local medium.

To demonstrate this through explicit calculation, we begin with the following equations from \cite{Deng2009}, ignoring terms of order ${\cal O}(\epsilon^4)$ and higher:
\begin{align}
g_{00}&=-1+\epsilon^2N\tag{3}\label{eq:3},\\
g_{ij}&=\delta_{ij}+\epsilon^2H_{ij}\tag{5}\label{eq:5},\\
G&=G_0(1+\epsilon^2\overset{(2)}G)\tag{25}\label{eq:25},\\
m&=m_0(1+\epsilon^2\overset{(2)}m)\tag{29}\label{eq:29},\\
\omega&=\omega_0(1+\epsilon^2\overset{(2)}\omega)\tag{30}\label{eq:30},\\
R_{00}&=-\frac{1}{2}\epsilon^2\Box N\tag{A13}\label{eq:A13},\\
T_{00}&=\sigma-\epsilon^2(2N\sigma+\sigma_{kk})\tag{A14}\label{eq:A14},\\
T_{ij}&=\epsilon^2\sigma_{ij}\tag{A16}\label{eq:A16},\\
T&=-\sigma+\epsilon^2(N\sigma+2\sigma_{kk})\tag{A17}\label{eq:A17}
\end{align}
from which it follows that $\partial_\mu G\partial^\mu G\propto\epsilon^4$, $\partial_\mu m\partial^\mu m\propto\epsilon^4$, and $\partial_\mu\omega\partial^\mu\omega\propto\epsilon^4$ can all be ignored at the $\epsilon^2$ level. Terms quadratic in $B_{\mu\nu}$ can also be ignored, leading us to rewrite Eq. (\ref{eq:12}) from \cite{Deng2009} as follows:
\begin{align}
R_{\mu\nu}=&\frac{8\pi G}{c^2}\left[T_{\mu\nu}-\frac{1}{2}g_{\mu\nu}T+\omega\frac{m^2k^2}{c^2}\phi_\mu\phi_\nu\right]\nonumber\\
&-\frac{G_{;\mu\nu}}{G}-\frac{1}{2}g_{\mu\nu}\frac{G_{;\kappa}^{;\kappa}}{G}.\tag{12}\label{eq:12}
\end{align}
The $00$-component of (\ref{eq:12}) can be written, using (\ref{eq:3}), (\ref{eq:A14}), and (\ref{eq:A17}), as
\begin{align}
R_{00}=&\frac{8\pi G}{c^2}\bigg[\sigma-\epsilon^2(2N\sigma+\sigma_{kk})\nonumber\\
&\hskip 24pt+\frac{1}{2}(1-\epsilon^2N)\left\{-\sigma+\epsilon^2(N\sigma+2\sigma_{kk})\right\}\nonumber\\
&\hskip 24pt+\omega\frac{m^2k^2}{c^2}\phi_0\phi_0\bigg]+\frac{1}{2}(1-\epsilon^2N)\frac{\Box G}{G},\tag{12a}
\end{align}
or, after simplification, while making use of (\ref{eq:25}) and (\ref{eq:A13}), we obtain a modified version of (\ref{eq:A18}) of \cite{Deng2009}:
\begin{equation}
-\frac{1}{2}\epsilon^2\Box N=\frac{8\pi G}{c^2}\left[\frac{1}{2}\sigma-\epsilon^2N\sigma+\omega\frac{m^2k^2}{c^2}\phi_0\phi_0\right]+\frac{1}{2}\epsilon^2\Box\overset{(2)}G.\tag{A18}\label{eq:A18}
\end{equation}

In a similar vein, we obtain the $ij$-component of (\ref{eq:12}). First, we make use of the gauge condition (32) in \cite{Deng2009} and rewrite (\ref{eq:A11}) from \cite{Deng2009} as
\begin{equation}
R_{ij}=\epsilon^2\left(-\frac{1}{2}H_{ij,kk}-\overset{(2)}G_{,ij}\right).\tag{A11}\label{eq:A11}
\end{equation}
Using (\ref{eq:A11}), we then write the $ij$-component of (\ref{eq:12}) as
\begin{align}
&\epsilon^2\left(-\frac{1}{2}H_{ij,kk}-\overset{(2)}G_{,ij}\right)=\frac{8\pi G}{c^2}\bigg[\epsilon^2\sigma_{ij}\nonumber\\
&-\frac{1}{2}(\delta_{ij}+\epsilon^2H_{ij})\{-\sigma+\epsilon^2(N\sigma+2\sigma_{kk})\}+\omega\frac{m^2k^2}{c^2}\phi_i\phi_j\bigg]\nonumber\\
&-\frac{G_{;ij}}{G}-\frac{1}{2}(\delta_{ij}+\epsilon^2H_{ij})\frac{G^{;\kappa}_{;\kappa}}{G}.\tag{12b}
\end{align}
After simplification, we obtain (\ref{eq:A20}) from \cite{Deng2009}, again with slight modifications:
\begin{align}
-\frac{1}{2}\epsilon^2H_{ij,kk}&=\frac{8\pi G}{c^2}\bigg[\frac{1}{2}\delta_{ij}\sigma+\epsilon^2\bigg(\sigma_{ij}-\frac{1}{2}\delta_{ij}N\sigma-\delta_{ij}\sigma_{kk}\nonumber\\
-\frac{1}{2}&\sigma H_{ij}\bigg)+\omega\frac{m^2k^2}{c^2}\phi_i\phi_j\bigg]-\frac{1}{2}\epsilon^2\delta_{ij}\Box\overset{(2)}G.\tag{A20}\label{eq:A20}
\end{align}

Next, we rewrite Eq.~(\ref{eq:20}) from \cite{Deng2009} as follows:
\begin{align}
&(\theta+3)\frac{G_{;\kappa}^{;\kappa}}{G}=\nonumber\\
&-\frac{8\pi G}{c^2}\bigg(-\sigma+\epsilon^2(N\sigma+2\sigma_{kk})-\omega\frac{m^2k^2}{c^2}\phi_\mu\phi^\mu\bigg),\tag{20}\label{eq:20}
\end{align}
from which
\begin{equation}
\epsilon^2\Box\overset{(2)}G=\frac{1}{\theta+3}\frac{8\pi G}{c^2}\bigg(\sigma-\epsilon^2(N\sigma+2\sigma_{kk})+\omega\frac{m^2k^2}{c^2}\phi_\mu\phi^\mu\bigg).\tag{20a}\label{eq:20a}
\end{equation}

Using (\ref{eq:20a}), we can rewrite (\ref{eq:A18}) as
\begin{align}
-\frac{1}{2}\epsilon^2\Box N=&\frac{8\pi G}{c^2}\bigg[\left(\frac{1}{2}+\frac{1}{2(\theta+3)}\right)\sigma+\omega\frac{m^2k^2}{c^2}\phi_0\phi_0\nonumber\\
&+\frac{1}{2(\theta+3)}\omega\frac{m^2k^2}{c^2}\phi_\mu\phi^\mu+{\cal O}(\epsilon^2)\bigg],\tag{A18a}
\end{align}
and (\ref{eq:A20}) as
\begin{align}
-\frac{1}{2}\epsilon^2H_{ij,kk}=&\frac{8\pi G}{c^2}\bigg[\left(\frac{1}{2}-\frac{1}{2(\theta+3)}\right)\delta_{ij}\sigma+\omega\frac{m^2k^2}{c^2}\phi_i\phi_j\nonumber\\
-\frac{1}{2(\theta+3)}&\delta_{ij}\omega\frac{m^2k^2}{c^2}\phi_\mu\phi^\mu+{\cal O}(\epsilon^2)\bigg].\tag{A20a}
\end{align}

As in \cite{Deng2009}, we define $H_{ij}=\delta_{ij}V$. If we ignore terms containing $\phi_i$, we get
\begin{equation}
\epsilon^2\Box N=-\frac{8\pi G}{c^2}\left[\frac{\theta+4}{\theta+3}\sigma+\frac{2\theta+5}{\theta+3}\omega\frac{m^2k^2}{c^2}\phi_0^2+{\cal O}(\epsilon^2)\right],\tag{A22}
\end{equation}
and
\begin{equation}
\epsilon^2\Box V=-\frac{8\pi G}{c^2}\left[\frac{\theta+2}{\theta+3}\sigma+\frac{1}{\theta+3}\omega\frac{m^2k^2}{c^2}\phi_0^2+{\cal O}(\epsilon^2)\right].\tag{A24}
\end{equation}
If $\sigma\gg\omega m^2k^2\phi_0^2/c^2$, we get back the result of \cite{Deng2009} for the Eddington--Robertson parameter $\gamma$:
\begin{equation}
\gamma=\frac{V}{N}\simeq\frac{\Box V}{\Box N}\simeq\frac{\theta+2}{\theta+4}.\tag{A25}\label{eq:A25}
\end{equation}
However, if $\phi_0^2$ is the dominant term, we obtain
\begin{align}
\gamma\simeq\frac{1}{2\theta+5}.\tag{A25a}\label{eq:A25a}
\end{align}
%and a value of $\theta=-2$ yields $\gamma=1$, consistent with solar system observations.
%
%In the near-perfect vacuum of the outer solar system, the matter stress-energy tensor is a vanishing quantity. In contrast, $\phi_0$ is nonvanishing \cite{Moffat2007e}:
The quantity $\phi_0$ is nonvanishing \cite{Moffat2007e}:
\begin{equation}
\phi_0\simeq -\sqrt{\frac{G_N}{\omega}}M\frac{e^{-mr}}{r}\nonumber
\end{equation}
is determined by the magnitude of the source mass $M$ (in this case, the mass of the Sun) and the inverse of the distance from the source. This expression allows us to calculate numerically the mass density associated with the vector field at a specific distance from the Sun, for instance at $r=0.1$~A.U.:
\begin{equation}
%\rho_\phi{}_{[r=0.1~\mathrm{A.U.}]}=\omega\frac{m^2k^2}{c^2}\frac{G_N}{\omega}M^2\frac{e^{-2mr}}{r^2}=5.4\times 10^{-19}~\mathrm{kg}/\mathrm{m}^3.
\rho_\phi{}_{[r=0.1~\mathrm{A.U.}]}=\frac{m^2k^2G_NM^2e^{-2mr}}{c^2r^2}=5.4\times 10^{-19}~\mathrm{kg}/\mathrm{m}^3.\nonumber
\end{equation}

In contrast, a typical proton density of 5 protons per cubic centimeter at a distance of 1~A.U. from the Sun (and assuming that the density of the solar wind falls off as $1/r^2$) we get, at $r=0.1$~A.U., given the proton mass at $m_p=1.67\times 10^{-27}$~kg, a proton mass density of
\begin{equation}
\rho_p{}_{[r=0.1~\mathrm{A.U.}]}=100m_pN_p{}_{[r=1~\mathrm{A.U.}]}=8.35\times 10^{-19}~\mathrm{kg}/\mathrm{m}^3.\nonumber
\end{equation}

Given that these two values, $\rho_\phi$ and $\rho_p$, are comparable in magnitude, neither (\ref{eq:A25}) nor (\ref{eq:A25a}) properly represent the observed value of the Eddington-parameter $\gamma$. Further, one must consider significant fluctuations (up to an order of magnitude) in the solar wind mass density, making it likely that no constant value of $\theta$ can satisfy the stringent limits on $\gamma$ from precision solar system observations.

% Therefore, $\phi_0$ will dominate, yielding (\ref{eq:A25a}).

%The approximate solution presented in \cite{Moffat2007e} is consistent with either choice of $\theta=-1$ (which corresponds to the Lagrangian used in \cite{Moffat2007e}) or $\theta=-2$, in accordance with (\ref{eq:A25a}).

\bibliography{refs}

\end{document}